\documentclass[prb,aps,twocolumn,floatfix]{revtex4}
\usepackage{graphicx}
\usepackage{epsfig}
\usepackage{amsmath}
\def \ds {\displaystyle}

\def \r12 {{\bf r}_{12}}
\def \ds {\displaystyle}

\def \HZJ {H^{\rm ZJ}}
\def \Dsw {\Delta_{{\rm SW}}}
\def \hm {{\bf m}}

\def \Tc {T_{\rm C}}

\def \vS {{\bf S}}

\def \Jc {J_c}
\def \vJ {{\bf J}}
\def \vk {{\bf k}}
\def \hk {\hat {\bf k}}
\def \kperp {{\bf k}_{\perp }}
\def \vq {{\bf q}}
\def \vj {{\bf j}}
\def \vR {{\bf R}}
\def \dg {\dagger}
\def \bk {{\bf k}}

\begin{document}

\date{\today}
\title{A Single Layer of Mn in a GaAs Quantum Well:  \\ a Ferromagnet with Quantum Fluctuations}
\author{Roger G. Melko, Randy S. Fishman, and Fernando A. Reboredo}
\affiliation{Materials Science and Technology Division, Oak Ridge National Laboratory, 
Oak Ridge, Tennessee 37831-6032}

\begin{abstract}

Some of the highest transition temperatures achieved for Mn-doped GaAs have been in
$\delta $-doped heterostructures with well-separated planes of Mn.  
But in the absence of magnetic anisotropy, the Mermin-Wagner theorem implies that a single plane of magnetic ions cannot be ferromagnetic.  
Using a Heisenberg model, we show that the same mechanism that produces
magnetic frustration and suppresses the transition temperature in bulk Mn-doped GaAs, 
due to the difference between the light and heavy band masses, can stabilize ferromagnetic 
order for a single layer of Mn in a GaAs quantum well.  This comes at the price of quantum 
fluctuations that suppress the ordered moment from that of a fully saturated ferromagnet.
By comparing the predictions of Heisenberg and Kohn-Luttinger models, we conclude that 
the Heisenberg description of a Mn-doped GaAs quantum well breaks down when the
Mn concentration becomes large, but works quite well in the weak-coupling limit of small Mn 
concentrations.   This comparison allows us to estimate the size of the quantum fluctuations in 
the quantum well.
  
\end{abstract}
\maketitle

\newpage

\section{Introduction}

The discovery of dilute-magnetic semiconductors (DMS) with transition temperatures above 
170 K  \cite{Ohno96,Mun89,MacDon05} has renewed hopes for a revolution in 
semiconductor technologies based on electron spin \cite{Zutic04}.  However, the difficulty in 
producing DMS materials with ferromagnetic transition temperatures above room temperature has 
stalled the development of practical spintronic devices.  Some of the highest transition temperatures 
to date have been achieved in $\delta $-doped GaAs heterostructures,\cite{Naz05} where the Mn 
are confined to isolated planes.   The Curie temperature of digital heterostructures does not seem to 
vanish as the distance between layers increases \cite{Kaw00}.
Yet a single layer of magnetic ions with spin $S=5/2$ would not be expected to become ferromagnetic
due to the Mermin-Wagner theorem, which states that gapless spin excitations destroy magnetic
order at any nonzero temperature in two dimensions.   The magnetic anisotropy due to strain 
is commonly invoked \cite{Lee00,Abol01,Liu03,Prio05} to explain the ferromagnetism of heterostructures.   
However, strain changes significantly as a function of film thickness and capping layers, and would be 
negligible for a single layer of Mn in a GaAs quantum well.  The questions addressed in this paper are:  
can a single layer of Mn in a GaAs quantum well be ferromagnetic and what kind of ferromagnet is it?

To answer these questions, we employ two complementary approaches.  First, we study the magnetic 
interactions between Mn moments embedded in a GaAs quantum well using a two-dimensional (2D) 
$S=5/2$ Heisenberg model, with exchange interactions taking the same form as in bulk Mn-doped GaAs.\cite{Zar02}   
Remarkably, the same anisotropic Heisenberg interactions that suppress the bulk transition temperature act to  
stabilize long-range magnetic order in 2D by producing a gap in the spin-wave (SW) spectrum.  
Second, we estimate the parameters of the Heisenberg model by constructing a Kohn-Luttinger (KL) 
model for a GaAs quantum well with an additional exchange interaction between the holes and
the Mn spins, which are treated classically and confined to the central plane.  Comparing the predictions 
of the Heisenberg and KL plus exchange (KLE) models, we conclude that the Heisenberg description of 
a Mn-doped quantum well works quite well for small Mn concentrations or in the weak-coupling limit,
but breaks down when the Mn concentration becomes too large

This paper is organized into five sections.   In Section~\ref{secZJ}, we examine several 
anisotropic $S=5/2$ Heisenberg models that might describe a single layer of 
Mn in a GaAs quantum well.  Section~\ref{secKLE} develops a more precise electronic description of the 
quantum well based on the KLE model.   In Section~\ref{secPAR}, the KLE model is used to 
estimate the parameters of an anisotropic Heisenberg model.   A brief conclusion is provided in 
Section~\ref{secCon}.

\section{Heisenberg description of a Quantum Well}
\label{secZJ}

In bulk Mn-doped GaAs, the difference between the light ($m_l=0.07m$) and heavy
($m_h=0.5m$) band masses with ratio $r=m_l/m_h\approx 0.14$ produces an anisotropic 
interaction \cite{Zar02} between any two $S=5/2$ Mn ions.  This anisotropy arises because the 
kinetic energy $K=\sum_{\vk, \alpha \beta }\epsilon (\vk )_{\alpha \beta }c_{\vk ,\alpha}^{\dagger }
c_{\vk \beta }$ is only diagonalized when the angular momentum $\vj $ of the charge carriers is quantized along 
the momentum direction $\hk $.   The heavy and light holes carry angular momentum $\vj \cdot \hk =\pm 3/2$ 
and $\pm 1/2$, respectively.  As shown by Zar\'and and Jank\'o,\cite{Zar02} the interaction between 
Mn spins $\vS_i$ and $\vS_j$ (as in the inset to Fig.~1) can be written
\begin{equation}
H_{ij}=-J_{ij}^{(1)} \vS_i \cdot \vS_j +J_{ij}^{(2)} \vS_i\cdot {\bf r}_{ij} \, \vS_j \cdot {\bf r}_{ij} .
\label{eqZJ}
\end{equation}
For $r=1$, $K$ is diagonal in any basis and $J_{ij}^{(2)} =0$.
Since $J_{ij}^{(2)} > 0 $ for $r< 1$, the Mn spins in GaAs prefer to align perpendicular to the 
vector ${\bf r}_{ij} =(\vR_i-\vR_j)/|\vR_i-\vR_j|$ connecting spins $i$ and $j$.
For a tetrahedron of Mn spins, the interaction energy between every pair of spins cannot be 
simultaneously minimized and the system is magnetically frustrated.
As shown both in the weak-coupling, RKKY limit \cite{Zar02} and more generally \cite{Ary05,Mor06} within
dynamical mean-field theory, this anisotropic interaction suppresses the Curie temperature 
compared to a non-chiral system with $r=1$.    Moreno {\em et al.} \cite{Mor06}
found that the transition temperature may be lowered by about 50\% for $r=0.14$ compared to $r=1$.

We construct a simple 2D Heisenberg model for a single layer of Mn-doped GaAs by including the effect of this 
anisotropy:
\begin{equation}
H^{ZJ} =\frac{1}{2}\sum_{i \ne j}H_{ij} - D \sum_i (S_i^z)^2
\label{Hzj}
\end{equation}
where the sum is taken over a square lattice and $S=5/2$.
We also include a single-ion anisotropy term $D$ that might be important in a quantum well due
to elastic strain and spin-orbit coupling.
When $J_{ij}^{(2)}>0$, the anisotropic interactions cause the Mn spins to point out of the plane, 
producing a gap $\Dsw $ in the SW spectrum even when $D=0$.
By breaking the rotational invariance of the spin, the anisotropic coupling would allow a single
layer of Mn ions to order ferromagnetically even in the absence of single-ion anisotropy.
As discussed further below, the single-ion anisotropy $D$ is also known \cite{Erick91} to produce a 
SW gap and to stabilize ferromagnetic long-range order in the 2D Heisenberg model.

Before proceeding to study $\HZJ $ using SW theory, we
note that other types of Heisenberg models may also stabilize ferromagnetic order in 2D.  For example, the Hamiltonian
\begin{equation}
H'=-J\sum_{\langle i,j\rangle} \Bigl\{ S_i^xS_j^x+S_i^yS_j^y+(1+\gamma )S_i^zS_j^z \Bigr\}
\end{equation}
with an anisotropic coupling between neighboring spins was recently used \cite{Prio05} to model a 
plane of Mn spins in a GaAs host.  When $\gamma \ne 0$, the rotational invariance of the spins is broken and 
2D ferromagnetism is stabilized at finite temperatures.    However, $H'$ does not contain the same anisotropic
interactions between Mn spins that are believed \cite{Zar02} to be present in the bulk system.

Due to the presence of the $J^{(2)}_{ij}$ terms in $H_{ij}$, $\HZJ $ does not commute with the total spin 
${\bf S}_{{\rm tot}}=\sum_i {\bf S}_i$.  Therefore, quantum fluctuations will suppress the magnetic moment of the
ZJ model even at zero temperature.   Such quantum fluctuations are typically found in antiferromagnets 
but are rather unusual in ferromagnets away from quantum critical points.  Notice that the Hamiltonian
$H'$ constructed above does not have quantum fluctuations.

To gain a better idea of the size of the quantum fluctuations in the ZJ model,
we specialize to the case where both $J^{(1)}_{ij}\equiv J_1$ and $J^{(2)}_{ij}\equiv J_2$ couple only
neighboring spins on the square lattice.  We then use SW theory to solve $\HZJ $, assuming that
$S \gg 1$.  At the mean-field ($1/S^0$) level, the ferromagnetic alignment of the spins along the $z$ axis is
unstable to an A-type antiferromagnetic realignment of the spins in the $xy$ plane 
(with lines of spins alternating in orientation) when $J_2/J_1 > 2+D/J_1$.   To order $1/S$, 
Eq.~(\ref{eqZJ})
may be written in terms of Holstein-Primakoff bosons:
\begin{eqnarray}
H_{ij} &\approx& -J_1 \Bigl\{ {S^2 
-S(a_i^{\dg}a_i + a_j^{\dg}a_j - a_i^{\dg}a_j- a_j^{\dg}a_i)  } \Bigr\} \nonumber \\
&\pm& \frac{J_2S}{2}(a_i \pm a_i^{\dg})(a_j \pm a_j^{\dg}),
\label{Hsw0}
\end{eqnarray}
where $\pm$ is $+$ for spin $i$ and $j$ separated by $\hat{x}$, and $-$ for spin $i$ and $j$ 
separated by $\hat{y}$.   The second term in Eq.~(\ref{Hzj}) will similarly contribute 
$2DS \sum_i a_i^{\dg}a_i$.  Since only the first terms in a $1/S$ expansion of the spin operators 
have been retained, the interactions between SW's have been neglected.  
Because of the relatively large $S=5/2$ spin of the magnetic ions, however, the next order in the 
expansion ($1/S^2$) will be rather small for low temperatures, and we expect the 
linear approximation to recover all of the relevant quantum physics.
Writing the Holstein-Primakoff bosons in a
momentum representation, we get the useful form:
\begin{eqnarray}
\HZJ &=& H_0 + 
\sum_{\bk} A_{\bk} \left({ a^{\dg}_{\bk}a_{\bk} + a^{\dg}_{-\bk}a_{-\bk}  }\right) 
\nonumber \\
&+& \sum_{\bk} B_{\bk} \left({ a^{\dg}_{\bk}a^{\dg}_{-\bk} + a_{\bk}a_{-\bk}  }\right),
\label{LSWham}
\end{eqnarray}
with coefficients given by
\begin{eqnarray}
A_{\bk} &=& 2J_1S \left[{ 1-\xi(\bk) }\right]
+ J_2S \xi(\bk) +DS  \nonumber \\
B_{\bk} &=& J_2S \phi(\bk)/2,
\end{eqnarray}
where $\xi(\bk) = (\cos{k_x} + \cos{k_y})/2$
and $\phi(\bk) = \cos{k_x}-\cos{k_y}$ (lattice constant set to unity).
In this form, it is easy to see that the anisotropy $J_2$
destroys the ability of the Hamiltonian to commute with the total spin 
$S_{{\rm tot}}^z=\sum_i S^z_i=NS -\sum_{\bk }a^{\dg }_{\bk}a_{\bk}$, 
causing a quantum correction to the ground state magnetization.

\begin{figure}[floatfix]
{
\includegraphics[width=3.3in]{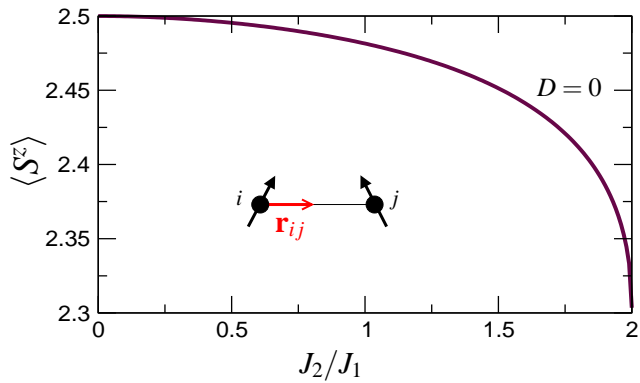}
\caption{(color online)  Quantum correction to the ground state magnetization per spin 
($S=5/2$) in the limit $D=0$.  The inset illustrates the direction of the interaction $J_{ij}^{(2)}$.}
}
\end{figure}

The SW Hamiltonian can be diagonalized in the usual way by transforming to Bogoliubov
bosons $\alpha^{\dagger}_{\bf k}$  and $\alpha_{\bf k} $, which obey the canonical commutation
relations.  Forcing the resulting Hamiltonian to 
be diagonal in $\alpha^{\dagger}_{\bf k} \alpha_{\bf k}$, we obtain the SW energies
$\omega_{\bf k} = 2\sqrt{A^2_{\bf k} - B^2_{\bf k}}$ with the energy gap $\Dsw = 2(J_2+D)S$.   
The absence of terms linear in the boson operators and the positive definiteness of the SW frequencies for 
$-D < J_2 < 2J_1+D$ guarantee that the ferromagnetic state with all spins aligned  
along the $z$ axis is stable against non-collinear rearrangements of the spins \cite{Sch02}.
The diagonalized SW Hamiltonian may be used to estimate the quantum-mechanical correction to 
the ground-state magnetization.  Starting with the magnetization written as
$
\langle S_i^z \rangle = S - N^{-1} \sum_{\bk} 
\langle{ a^{\dg}_{\bk} a_{\bk} }\rangle,
$
we find upon transforming to Bogoliubov bosons, 
\begin{equation}
  \langle S_i^z \rangle =   S 
- \frac{1}{N} \sum_{\bf k} \biggl\{ {
  \frac{A_{\bf k}}{\omega_{\bf k} } - \frac{1}{2}
 +  \frac{2A_{\bf k}}{\omega_{\bf k} }  \langle \alpha^{\dagger}_{\bf k} \alpha_{\bf k} \rangle
}\biggr\} .
\label{Sz2}
\end{equation}
Since 
$ n_{\bf k}  = \langle \alpha^{\dagger}_{\bf k} \alpha_{\bf k} \rangle =1/( {\textrm e}^{\beta \omega_{\bf k}} -1 ),$
Eq.~(\ref{Sz2}) can be used to calculate the quantum correction to the ground-state magnetization by 
simply setting $\langle \alpha^{\dagger}_{\bf k} \alpha_{\bf k} \rangle=0$ and evaluating the integrals over ${\bf k}$. 
The result is plotted versus $J_2 / J_1$ for $D=0$ in Fig.~1.

\begin{figure}[floatfix]
{
\includegraphics[width=3.1in]{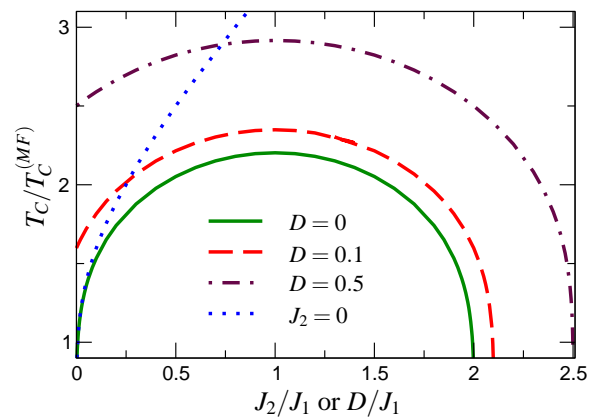}
\caption{(color online)  The transition temperature of the nearest-neighbor $H^{ZJ}$ model 
calculated in SW theory.  For $D=0$, $T_C$ falls to zero at $J_2/J_1=0$ and 2 ;  for $J_2=0$, $T_C=0$ for 
$D=0$ only.  The SW value for $\Tc $ is normalized by the Weiss MF result.
\label{2DTc}}
}
\end{figure}

Within linear SW theory, Eq.~(\ref{Sz2}) also affords us the simplest estimate for the transition 
temperature $\Tc$ of the nearest-neighbor $\HZJ$ model.  Clearly, the transition temperature calculated 
within this formalism neglects the large quantitative corrections due to SW interactions.  But  
including such interactions to second-order in $1/S$ would require additional
approximations to handle the terms quartic in $\alpha^{\dagger}_{\bf k}$ and $\alpha_{\bf k}$.
Since the precise value of $\Tc $ does not play a role in the subsequent
discussion, we are satisfied that the correct qualitative trends (i.e. the
stabilization of $\Tc $ at a finite value) are reproduced by linear SW theory.
Including now both terms of Eq.~(\ref{Sz2}), we can write the $T$-dependent 
order parameter as 
\begin{equation}
\langle S_i^z \rangle = S+\frac{1}{2} 
- \frac{1}{N} \sum_{\bf k} {  \frac{A_{\bf k}}{\omega_{\bf k}  }
 \coth \left({ \frac{\beta \omega_{\bf k}  }{2} }\right)  } .
\end{equation}
We obtain $\Tc $ from the crossing point
$\langle S_i^z \rangle = 0$ for a given set of parameters $J_2/J_1$ or $D/J_1$ \cite{finiteT}.  
The result is illustrated in Fig.~\ref{2DTc}, where $\Tc $ is scaled by the Weiss mean-field (MF) result,
$\Tc^{(MF)} = 4J_1 S(S+1)/3$.

Consider first the limiting case with $D=0$.
Because the SW stiffness is assumed to be independent of temperature, the SW approach 
overestimates $\Tc $ for $J_2/J_1\approx 1$.   Notice that $\Tc $ is symmetric on either side of 
$J_2/J_1=1$ and drops to zero when $J_2/J_1=0$ and 2.  The limit $J_2/J_1 \rightarrow 0$
can be analyzed by considering the behavior of
the SW frequencies for small ${\bf k}$.  Expanding $\omega_{\bf k}$ to second order in  ${\bf k}$, 
we find that for small $J_2/J_1$, 
$\Tc  \rightarrow 4 \pi J_1S(S+1/2)/ \ln (4 \pi^2 J_1/J_2)$.  
This inverse logarithm is similar to the 
form discussed in Ref.~[\onlinecite{Erick91}] for the $\HZJ$ model with $J_2=0$ and small single-ion 
anisotropy.  In that case, $\Tc  \rightarrow T_3 / \ln ( \pi^2 J_1/D)$, where $T_3$ is the bulk transition 
temperature, approximately proportional to $J_1 S(S+1)$.

We confirm this relationship by calculating $T_C$ in the $\HZJ$ model using the above SW theory, 
with $J_2=0$ and a nonzero $D$.   As shown in Fig.~\ref{2DTc}, the results for $J_2=0$ or $D=0$ agree 
quite well as $D$ or $J_2\rightarrow 0$.  For larger $J_2$ and $D$, the curves deviate significantly, with
the $J_2=0$ curve growing linearly in the limit of large $D$.   Quite generally, we find that as 
$\Dsw \rightarrow 0$,
\begin{equation}
\Tc  \rightarrow \frac{4 \pi J_1S \left(S+1/2 \right)}
{ \ln \left( 8 \pi^2 J_1S/\Dsw \right)}.
\end{equation}
This general expression reduces to the correct limits when $J_2$ or $D=0$.  
We conclude that the behavior of $\Tc$ in the case of strong fluctuations is controlled
by the isotropic exchange $J_1$ and the SW gap.

With a finite $D$, the $\Tc $ versus $J_2/J_1$ curve qualitatively resembles a scaled-up version of the $D=0$ 
curve in Fig.~\ref{2DTc}.  In that case however, the asymptotic behavior at $J_2=0$ is removed ($\Tc $ becomes finite), 
the instability to in-plane ordering occurs at $J_2/J_1 = 2+D/J_1$ (as discussed above), and the maximum $\Tc $ 
increases.   In our SW calculation, $\Tc $ vanishes at this instability due to the softening of the SW frequencies 
$\omega_{\bf k}$ with ${\bf k}=(\pi, 0)$ and ${\bf k}=(0,\pi )$.

\section{The Kohn-Luttinger plus exchange model}
\label{secKLE}

To estimate the size of the quantum fluctuations in a Mn-doped GaAs quantum 
well, we model the quasi-2D hole gas as a thin quantum well of width $L$ accounting for the
Coulomb confinement potential arising from the ionized
dopants \cite{reboredo93}, central cell corrections, and any additional epitaxial confinement.  
We use a spherical approximation of the KL Hamiltonian \cite{Bald73} with four bands 
($\vj \cdot \hk =\pm 3/2$ and $\pm 1/2$) to evaluate the energies $\epsilon (\vk )_{\alpha  \beta }$ 
  for a GaAs quantum well with zero boundary conditions at
$z=\pm L/2$.  We express the Kohn-Luttinger Hamiltonian in a reduced basis
form by the lowest-energy wavefunctions 
$\psi_1(z)=\sqrt{2/L}\cos(\pi z/L)$ and $\psi_2(z)=\sqrt{2/L} \sin
(2\pi z/L)$ of the quantum well, with $\langle n| k_z^2| m \rangle =
(n\pi /L)^2\delta_{nm}$.  The Mn spins are now treated classically and
the Mn impurities are distributed in the $z=0$ plane with concentration $c$.
Since $\psi_2(0)=0$, the Mn
spins only couple to the holes in the first wavefunction with projection 
$\vert \psi_1(0)\vert^2 = 2/L $.  The exchange coupling of the Mn spins
$\vS_i=S\hm_i$ with the holes is given by $V=-2J_c\sum_i \hm_i \cdot
\vj_i $, where $\vj_i=\sum_{\alpha \beta }c_{i,\alpha }^{\dagger
}\vJ_{\alpha \beta }c_{i,\beta }/2$ are the hole spins and
$\vJ_{\alpha \beta}$ are the Pauli spin-3/2 matrices.  The states $\psi_1 $
and $\psi_2$ are coupled by the off-diagonal terms in the KL
Hamiltonian with matrix elements proportional to $\langle n|(k_x\pm
ik_y)k_z|m\rangle $, which vanishes for $n=m$ but is given by
$-(8i/3L)(k_x\pm i k_y)$ for $n=1$ and $m=2$.

For any coupling constant $J_c$ and orientation $\hm =(\sin\theta ,0 ,\cos \theta )$ of the Mn spins, 
the energy $E(J_c, \theta )$ of the KLE model is obtained by first diagonalizing 
an $8 \times 8$ matrix in $j=3/2$ and $n,m=1,2$ space:
\begin{equation}
 H^{KLE} = \left( \begin{array}{cc}
A_1 & B^{\dagger}  \\
B & A_2 \end{array} \right) . 
\end{equation}
The diagonal block elements can be written in terms of a generalized $4 \times 4$ matrix, 
\begin{widetext}
\begin{equation}
A_n^0=\left( \begin{array}{cccc}
\ds\frac{ k^2_{\perp}}{2m_a}+ \left({\ds\frac{n \pi}{L} }\right)^2 \ds\frac{1}{2m_h} +bQ_{\epsilon} 
        & 0 & d( k_x-i k_y)^2 & 0\\
0 & \ds\frac{ k^2_{\perp}}{2m_b}+ \left({\ds\frac{n \pi}{L} }\right)^2 \ds\frac{1}{2m_l} -bQ_{\epsilon} 
         & 0 & d( k_x-i k_y)^2 \\
d( k_x+i k_y)^2 & 0 & \ds\frac{ k^2_{\perp}}{2m_b}+ \left({\ds\frac{n \pi}{L} }\right)^2 \ds\frac{1}{2m_l} -bQ_{\epsilon}& 0 \\
0 & d( k_x+i k_y)^2 & 0 & \ds\frac{ k^2_{\perp}}{2m_a}+ \left({\ds\frac{n \pi}{L} }\right)^2 \ds\frac{1}{2m_h} +bQ_{\epsilon}
\end{array} \right) , 
\end{equation}so that $A_1=A_1^0-J_c{\bf m}\cdot{\bf J}$.   We include the strain term 
$Q_{\epsilon} \equiv \epsilon_{zz}-(\epsilon_{xx}+\epsilon_{yy})/2$, which is
multiplied by the deformation potential $b=-1.7$ eV for GaAs \cite{Dietl01}.
The other block diagonal element is simply $A_2=A_2^0$, since the holes in
$\psi_2(z)$ do not couple to the Mn spins at $z=0$.
We have defined $d=\sqrt{3}(1/m_h-1/m_l)/8$, $\kperp =(k_x, k_y)$,
$m_a=4(3/m_l+1/m_h)^{-1}$ and $m_b=4(1/m_l+3/m_h)^{-1}$.
The off-diagonal block elements coupling $\psi_1(z)$ and $\psi_2(z)$ are
\begin{equation}
B=\left( \begin{array}{cccc}
0 & {16id( k_x-i k_y)}/{3L}  & 0 & 0 \\
 {16id( k_x+i k_y)}/{3L} & 0 & 0 & 0 \\
0 & 0 & 0 & {-16id( k_x-i k_y)}/{3L} \\
0 & 0 & {-16id( k_x+i k_y)}/{3L} & 0
\end{array} \right) ,
\end{equation}
\end{widetext}
which couples the $j_z=\pm 3/2$ and $\pm 1/2$ components.

An expression for $J_c$ can be derived by directly comparing the potential $V$ to the standard 
spin-hole interaction, e.g.~Eq.~(1) of Ref.~[\onlinecite{Dietl01}].  We estimate 
\begin{equation}
\label{Jc}
J_c\approx \ds\frac{S(\beta N_0)}{2j\Lambda }c,
\end{equation}
where $S=5/2$, $j=3/2$, $\beta N_0 \approx 1.2$ eV,\cite{Dietl01}, $\Lambda =L/a$ is
the number of layers in the quantum well, and $a\approx 4$ $\overset{\circ}{\rm A}$ is the lattice constant of Ga
in the $z=0$ plane.  So we see that $J_c \approx 1\, c \, {\rm eV} /\Lambda $
is inversely proportional to the width of the quantum well.  For $\Lambda=10$, $J_c\approx 100\, c$ meV.

The splitting between the light and heavy band masses at the $\Gamma $ point
of $\psi_n(z)$ is given by
\begin{equation}
\Delta_n=\frac{1}{2}\left(\ds\frac{n\pi}{\Lambda a}\right)^2 \left({ \frac{1}{m_l}-\frac{1}{m_h} }\right)-2bQ_{\epsilon}.
\end{equation}
Since both contributions are positive for $Q_{\epsilon}<0$, the confinement of holes in the 
quantum well formally plays the same role as the compressive
strain in a thin film.\cite{Saw04}
For $\Lambda < 20$ and typical strains less than 0.5\%, the 
contribution to the band splitting from quantum confinement is much larger than the contribution from strain.
In the limit $\Lambda \rightarrow 0$ of an infinitely narrow quantum well, the effects of strain can be 
neglected entirely.  For a narrow quantum well with $\Lambda < 20$,
$\Delta_n\approx 29 (n/\Lambda )^2$ eV and $J_c /\Delta_1\approx \Lambda c/29$.
Hence, a narrow quantum well with a small Mn concentration is in the weak-coupling limit
with $J_c \ll \Delta_1$.    By contrast, GaAs films with $\Lambda $ larger than about 50 cannot be 
treated as quantum wells because too many wavefunctions $\psi_n(z)$ would be required. 
The dominant contribution to the band splitting in such films comes from 
strain \cite{Abol01} rather than the confinement of the holes.

After obtaining the eigenvalues of the $8\times 8$ matrix $H^{KLE}$, we evaluate
the energy $E(J_c, \theta )$ of a quantum well by integrating over $\kperp $.  
Since the hole filling $p$ in the region of perpendicular anisotropy is quite small, 
the holes only occupy a very small portion of the Brillouin zone centered around $\kperp =0$.  
This implies that each hole interacts with many different Mn moments, so that
the precise geometry and location of the Mn impurities within the $z=0$ plane does not 
affect our results.   Consequently, the results
of the KLE model for the energy do not depend on whether the Mn ions are randomly 
distributed within the central plane.   The small area in momentum space occupied by the
holes also justifies our use of a spherical approximation \cite{Bald73} to the KL Hamiltonian.

Because each of the holes must interact with two Mn moments in order to mediate their 
effective interaction, the resulting quantum-well energy $E(J_c,\theta )$ is an 
even function of $J_c$.
Of course, magnetic properties like the Kondo effect do depend on the sign of the
exchange coupling.  But in the weak-coupling limit of small $\Jc $, the Kondo temperature
\cite{White07} will be extremely small and the Kondo effect shall be neglected in the
subsequent discussion.

\section{Estimation of the Heisenberg parameters}
\label{secPAR}

By comparing the predictions of the KLE and ZJ models, we now estimate the ZJ exchange
interactions between the Mn moments in a GaAs quantum well.
In general, obtaining the long-range interactions 
$J^{(1)}_{ij}$ and $J^{(2)}_{ij}$ that parameterize the ZJ Hamiltonian is not possible.
However, we {\it can} estimate the $\vq=0$ component of the isotropic exchange coupling,
$J_1(\vq =0)=\sum_j J^{(1)}_{ij}$,
by considering the change in energy
of the quantum well when the exchange with the Mn spins (all aligned along the $z$ direction with angle
$\theta =0$) is turned off.   For $\HZJ $, this gives the energy to align all of the spins.  So we find
that
\begin{equation}
\label{e1}
-\frac{1}{2}J_1(\vq =0)S^2-DS^2 =\frac{1}{N}\Bigl\{ E(J_c,\theta=0)-E(J_c=0)\Bigr\}.
\end{equation}
The $\vq=0$ component of the anisotropic exchange coupling, $J_2(\vq=0)=\sum_j J^{(2)}_{ij}$, may be
estimated by evaluating the change in energy when all of the Mn spins rigidly 
rotate away from the $z$ axis towards the $xy$ plane with angle $\theta $:
\begin{equation}
\label{e2}
\frac{1}{4}J_2 (\vq=0, \theta ) S^2 +DS^2 = \frac{1}{N}\frac{ E(J_c,\theta )-E(J_c,\theta=0)}{\sin^2 \theta }.
\end{equation}
Within the KLE model, the average exchange vanishes when the moments are randomly
oriented (whether perpendicular or parallel to the plane) because
holes with small momenta interact with many Mn moments.
By contrast, the single-ion anisotropy energy $-D \sum_i (S_i^z)^2$
is large when the moments are randomly oriented in the $z$ direction but vanishes
when the moments are randomly oriented in the plane.   Comparing the energies of these two
random configurations, we conclude that the single-ion anisotropy $D$ corresponding to the KLE model
must vanish.  Because the right-hand side of Eq.~(\ref{e2}) is independent of $\theta$ within the 
Hamiltonian Eq.~(\ref{Hzj}), the proximity of $J_2 (\vq=0,\theta \rightarrow 0)$ and $J_2 (\vq=0, \pi/2)$ 
is a measure of how well the KLE model can be approximated by the Heisenberg model $\HZJ $.

\begin{figure}
\includegraphics *[scale=0.42]{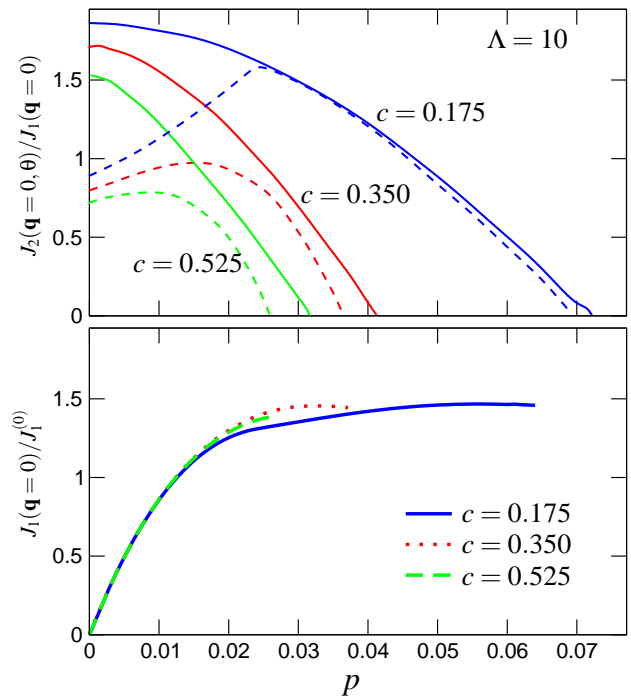}
\caption{(color online)
The top panel shows $J_2({\bf q}=0,\theta=0) /J_1({\bf q}=0)$ (dashed) and 
$J_2({\bf q}=0,\pi/2) /J_1({\bf q}=0) $ (solid) versus $p$ (holes per Mn) for three values of 
the Mn concentration $c$.   In the bottom panel, $J_1({\bf q}=0)/J_1^{(0)}$ is plotted versus
$p$ for the same concentrations.}
\end{figure}

Analytic results for the $\vq=0$ exchange parameters can be obtained in the 
weak-coupling limit of small $J_c$ if only $\psi_1(z)$ is occupied by holes and the 
hole chemical potential $\mu \approx \pi pc/m_a a^2$ is also small compared to the band splitting $\Delta_1$.   
For $J_c \ll \mu \ll \Delta_1$,  we find that $J_1(\vq=0) S^2=9\pi J_c^2 /W \equiv J_1^{(0)} S^2$ 
and $J_2(\vq=0,\theta ) = 2J_1(\vq=0)$, where $W = \pi^2c/m_a a^2$ is the bandwidth 
and $m_a$ is the band mass in the $xy$ plane.  Both $J_1(\vq=0)$ and 
$J_2(\vq=0,\theta )$ are independent of the hole filling due to the flatness of the 2D 
density-of-states \cite{Lee00}.  Recall from our discussion above
that the system with only nearest-neighbor interactions is unstable to an antiferromagnetic 
realignment of the spins in the $xy$ plane when $J_2/J_1> 2$.   So in the 
limit $\Jc \ll \mu \ll \Delta_1$, a phase with aligned moments is unstable when the exchange
interactions are short-ranged.   Neglecting the effect of 
anisotropy, $J_1(\vq=0)$ may be used to estimate $\Tc $ from the MF solution of a 
spin $S$ Heisenberg model: 
\begin{eqnarray}
\label{TCMF}
\Tc^{(MF)}&=& \frac{1}{3}J_1^{(0)} S(S+1) = \frac{1}{3\pi}(\beta N_0 )^2m_a a^2 S(S+1) \frac{c}{\Lambda^2}\cr &
\approx & 0.18\frac{c}{\Lambda^2}\, {\rm eV},
\end{eqnarray} 
which agrees precisely with the MF result of Lee {\em et al.} \cite{Lee00}   Hence,
the transition temperature of a quantum well increases as it becomes narrower.  
A departure of the concentration $c(z)$ of the Mn profile from a delta function would 
increase the effective width of the confining potential, thereby lowering the transition temperature.

More generally, when both wavefunctions $\psi_1(z)$ and $\psi_2(z)$ are included, 
Eqs.~(\ref{e1}) and (\ref{e2}) may be used 
to evaluate $J_2(\vq=0,\theta )/J_1(\vq=0)$ as a function of $p$.  As illustrated in Fig.~3, 
$J_2(\vq=0,0) /J_1(\vq=0)$ initially increases with filling before decreasing and becoming negative 
above $p_1$ (indicating that the spins with $\theta =0$ are no longer locally stable).  
We find that $J_1(\vq=0)$ increases linearly for small $p$ and reaches a value of about 1.4 times 
$J_1^{(0)}$ at a filling $p \approx 0.025$, independent of $c$.  The maximum in 
$J_2(\vq=0,0) /J_1(\vq=0)$ approaches 2 for small Mn concentrations $c$ and hole fillings $p$ 
approaching 0, which is just the limit $\Jc \ll \mu \ll \Delta_1$ discussed above.  By contrast, 
$J_2(\vq=0,\pi/2) /J_1(\vq=0)$ decreases with increasing $p$ and becomes negative when 
$p > p_2 > p_1$.  This implies that the moments are tilted away from 
the $z$ axis for $p > p_1$ and only fall into the $xy$ plane ($E(J_c,\theta )$ has a minimum 
at $\theta =\pi/2$) above some higher filling $p_3 > p_2$.   Once the spins land on 
the $xy$ plane above $p_3$, the rotational invariance of the spins about the $z$ axis would 
destroy long-range magnetic order if not for crystal-field anisotropy within the plane \cite{Saw04}.

As remarked earlier, the Heisenberg description of a quantum well requires that
$J_2(\vq =0,\theta )$ is independent of the angle $\theta $.   Hence, Fig.~3 implies that
for any nonzero Mn concentration, the ZJ model fails when the hole filling falls below a critical threshold.
The ZJ model works best over the widest range of hole fillings for small Mn concentrations.   
For $c=0.175$, the ZJ model is appropriate for hole fillings in the range $0.025 < p < 0.07$.    
As $c$ increases, the Heisenberg description only works within a narrower range of hole fillings, 
and even there not as well.  For Mn concentrations that are insufficiently small, the interactions 
between the Mn moments within the quantum well are too complex to be accounted for by the 
ZJ model, even when the exchange interactions are long-ranged.    

Up to this point, we have not made any assumptions about the range of the 
interactions $J^{(1)}_{ij}$ and $J^{(2)}_{ij}$ in the ZJ model.  When the Mn concentration is 
sufficiently low, we expect the nearest-neighbor interactions $J_1$ and $J_2$ to be the 
dominant ones and our results for $J_1(\vq=0)$ and $J_2(\vq=0,\theta )$ can be used to 
estimate the transition temperature of a quantum well.  In Section II, we found that the 
transition temperature reaches a peak when $J_2= J_1$.  So Fig.~3 suggests that the 
transition temperature of a quantum well will be highest for some hole filling slightly below
$p_1$, when the moments still lie along the $z$ axis and $J_1(\vq=0)\approx J_2(\vq=0,\theta )$.    
For $c=0.175$, $\Tc $ will be go through a maximum when $p\approx 0.048$ holes per Mn.

\section{Conclusions}
\label{secCon}

Our results imply that a single layer of Mn-doped GaAs (or a very thin film with $\delta $-doping)
with perpendicular magnetic moments may achieve a transition temperature close to the 
MF result of Eq.~(\ref{TCMF}).  Figure 2 suggest that $\Tc $ will reach a 
maximum with increasing $p$ when $J_2(\vq=0,\theta )\approx J_1(\vq=0)$.  
At higher fillings, $\Tc $ is expected to decrease until the Mn moments fall into the $xy$ plane, 
whereupon crystal field anisotropy is required to stabilize long-range ferromagnetic order.  

However, the description of a Mn-doped GaAs quantum well by a 
Heisenberg model, even one with long-range interactions, is restricted to 
small Mn concentrations and hole fillings that are above a threshold
value.   From Eq.~(\ref{Jc}), we see that this is just the
weak-coupling limit $\Jc \ll \Delta_1$ and $\Jc \ll \mu $.   As the Mn concentration increases, 
the Heisenberg description is valid over a narrower range of hole fillings and even within that 
range, is not as accurate.  Recent work \cite{Fis06} on the double-exchange model also 
reached the conclusion that a mapping onto a Heisenberg model is only valid in the weak-coupling limit.

For small Mn concentrations, where the Heisenberg description works quite well,
quantum fluctuations in the total spin may be at least partially responsible
for the depression of the magnetic moment found in Mn-doped GaAs epilayers.\cite{Pot02}
It is ironic that the same frustration mechanism that suppresses the Curie temperature 
and electronic polarization in bulk Mn-doped GaAs should permit ordering in a  
2D quantum well.   
With these results, we are in a position to address our originally postulated questions:
how can a single layer of Mn-doped GaAs be ferromagnetic, and what kind of ferromagnet is it?  
We conclude that a Mn-doped GaAs quantum well is ferromagnetic due to a SW gap produced by
the difference between the light and heavy band masses, but quantum fluctuations 
suppress the $T=0$ moment of this ferromagnet from its fully saturated value. 

It is a pleasure to acknowledge helpful conversations with Juana Moreno,
Thomas Maier, and Adrian Del Maestro.  This research was sponsored by the 
U.S. Department of Energy Division of Materials Science and Engineering under contract 
DE-AC05-00OR22725 with Oak Ridge National Laboratory, managed by UT-Battelle, LLC.

\end{document}